\documentclass[sigconf]{acmart}

\usepackage{booktabs}
\usepackage{multirow}
\usepackage{amsmath}

\AtBeginDocument{%
  }

\copyrightyear{2026}
\acmYear{2026}
\setcopyright{cc}
\setcctype{by}
\acmConference[AVI '26]{Proceedings of the 2026 International Conference on Advanced Visual Interfaces}{June 08--12, 2026}{Venice, Italy}
\acmBooktitle{Proceedings of the 2026 International Conference on Advanced Visual Interfaces (AVI '26), June 08--12, 2026, Venice, Italy}
\acmDOI{10.1145/3811427.3811450}
\acmISBN{979-8-4007-2342-1/2026/06}

\begin{document}

\title{LLM-Augmented Semantic Steering of Text Embedding Projection Spaces}


\author{Wei Liu}
\orcid{0009-0009-6340-8912}
\affiliation{%
  \institution{Virginia Tech}
  %\department{Department of Computer Science}
  \city{Blacksburg}
  \state{Virginia}
  \country{USA}
}
\email{wliu3@vt.edu}

\author{Eric Krokos}
\orcid{0000-0003-1350-5297}
\affiliation{%
  \institution{Department of Defense}
  \city{Washington}
  \state{DC}
  \country{USA}
}
\email{ericpkrokos@gmail.com}

\author{Kirsten Whitley}
\orcid{0000-0003-1356-326X}
\affiliation{%
  \institution{Department of Defense}
  \city{Washington}
  \state{DC}
  \country{USA}
}
\email{visual.tycho@gmail.com}

\author{Rebecca Faust}
\orcid{0000-0002-7640-1287}
\affiliation{%
  \institution{Tulane University}
  % \department{Department of Computer Science}
  \city{New Orleans}
  \state{Louisiana}
  \country{USA}
}
\email{rfaust1@tulane.edu}

\author{Chris North}
\orcid{0000-0002-8786-7103}
\affiliation{%
  \institution{Virginia Tech}
  % \department{Department of Computer Science}
  \city{Blacksburg}
  \state{Virginia}
  \country{USA}
}
\email{north@vt.edu}

\renewcommand{\shortauthors}{Liu et al.}

\begin{abstract}
Low-dimensional projections of text embeddings support visual analysis of document collections, but their spatial organization may not reflect the relationships an analyst intends to examine. Existing semantic interaction approaches encode semantic intent indirectly through geometric constraints or model updates, limiting interpretability and flexibility. We introduce LLM-augmented semantic steering, which enables analysts to express semantic intent by grouping a small set of example documents within the projection. A large language model externalizes this intent as natural-language representations and selectively extends it to related documents; the resulting semantic information is then incorporated into document representations via text augmentation or embedding-level blending, without retraining the underlying models. A case study illustrates how the same corpus can be reorganized from different semantic perspectives, while simulation-based evaluation shows that semantic steering improves global and local alignment with target semantic structures using only minimal interaction. Embedding-level blending further enables continuous and controllable steering of projection layouts. These results position projection spaces as intent-dependent semantic workspaces that can be reshaped through explicit, interpretable, language-mediated interaction.
\end{abstract}




\begin{teaserfigure}
  \centering
  \includegraphics[width=\textwidth]{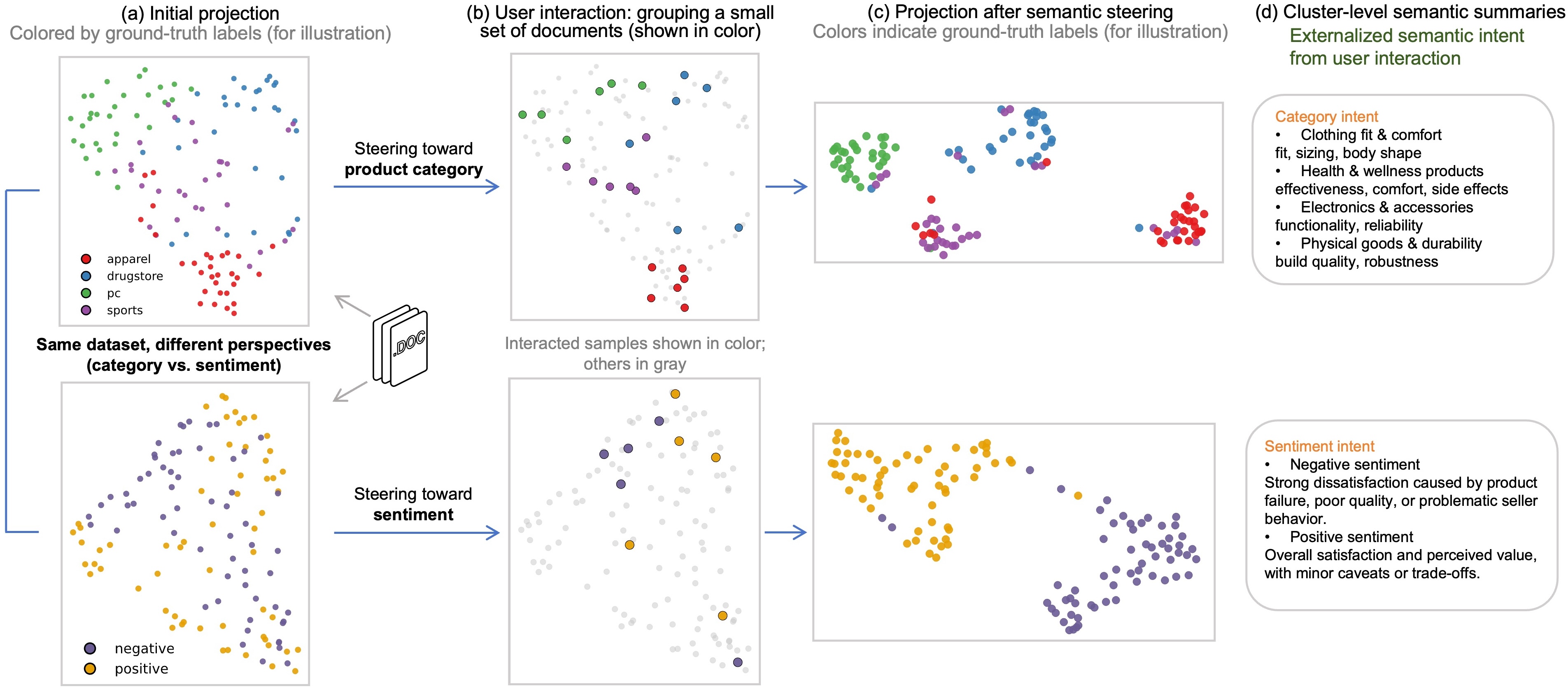}
  \caption{Semantic steering of projection spaces under different analytic perspectives. (a) Baseline projections generated under the same setup, shown from two perspectives: product category (top) and sentiment (bottom). (b) Analysts express semantic intent by grouping a small set of documents. (c) Updated projections after the semantic intent is externalized, selectively extended to related documents, and incorporated into document representations. (d) Cluster-level summaries generated by the LLM provide interpretable descriptions of the externalized semantic intent.}
  \Description{A two-by-four grid of scatterplots showing the same document collection under two analytic perspectives. The top row colors points by product category (apparel, drugstore, pc, sports) and the bottom row colors points by sentiment (negative, positive). Each row has four columns: an initial projection with intermixed points, a view where a small subset of grouped documents is highlighted in color while the rest are grayed out, an updated projection in which points of the same category or sentiment form more coherent regions, and a text panel showing cluster-level semantic summaries derived from the interaction.}
  \label{fig:teaser}
\end{teaserfigure}

\maketitle

\section{Introduction}
Low-dimensional projections of text embeddings are widely used in visual text analytics to support the exploration of document collections through spatial overviews \cite{jeon2025stop, huang2023va+, sacha2016visual}. By mapping high-dimensional representations into a two-dimensional (2D) space, these projections reveal clusters, neighborhoods, and outliers that facilitate hypothesis generation and pattern discovery \cite{jeon2025stop, liu2024visualizing}. 
In practice, however, projection layouts often fail to reflect an analyst's semantic intent. Documents that appear close in the projection may be conceptually unrelated for a given task, while documents that analysts perceive as semantically coherent may be scattered across the space. 
As a result, analysts often need to adapt their reasoning to the structure produced by the model, rather than reshaping the projection to reflect the semantic relationships they wish to examine.

Our goal is not to recover a single ``correct'' semantic organization of a dataset. Instead, we treat projections as \textbf{intent-dependent semantic workspaces}, where different analytic goals may legitimately give rise to different spatial organizations. From this perspective, interaction should enable analysts to express and refine semantic intent, allowing projection spaces to be reshaped to emphasize task-relevant dimensions \cite{endert2012semantics, sacha2016visual}.

Prior work has explored ways for analysts to influence projections through interaction, often by interpreting such interactions as geometric constraints or signals for updating model parameters \cite{bian2021deepsi, endert2012semantics, self2018observation, keith2023mixed, bian2019deepva, dowling2018bidirectional, sacha2016visual}. While effective in certain settings, these approaches encode semantic intent indirectly through geometric constraints and often require retraining or modifying embedding models, making them computationally expensive, model-dependent, and difficult to interpret---particularly for modern black-box embeddings. More recent approaches incorporate explicit semantic signals, such as labels or descriptors, into embedding and projection spaces, but typically rely on predefined schemas or dataset-wide annotation \cite{el2019semantic, oliveira2025creating}. These limitations highlight the need for mechanisms that allow analysts to express semantic intent directly in an interpretable and adaptable form.

Recent advances in large language models (LLMs) \cite{wu2025aligning, zhao2024lightva, kim2025intentflow, wang2025intentprism, gao2024taxonomy, wang2024survey} provide new opportunities to externalize semantic intent as natural language.
In this work, we introduce \textbf{LLM-augmented semantic steering} of text embedding projections. Our approach enables analysts to express semantic intent directly by grouping a small number of related documents within the projection.
An LLM externalizes the semantic intent inferred from these examples as structured representations and selectively extends it to related documents; the resulting semantic information is incorporated into document representations via augmentation or embedding-level blending.
This process updates the projection without retraining or modifying the embedding model or the dimensionality reduction (DR) method. Unlike approaches based on labeling or exhaustive assignment, semantic representations in our framework serve as intermediate language-based signals to guide representation updates, and are applied only when sufficient semantic evidence is available.

We demonstrate the effectiveness of this approach through both qualitative and quantitative evaluation. A case study on a product review corpus illustrates how the same dataset can be reorganized under different semantic perspectives, such as product category and sentiment. We further conduct simulation-based experiments in which interaction is controlled by sampling a small set of documents as example-based inputs, using target semantic groupings as references to evaluate alignment, interaction efficiency, and extension behavior. Results show that semantic steering improves alignment with minimal interaction and supports selective extension beyond documents directly involved in the interaction.

This work makes the following contributions:

\begin{itemize}
    \item A language-mediated approach for semantic steering of text embedding projections through lightweight interaction.
    
    \item An LLM-based mechanism that externalizes and selectively extends analyst-expressed semantic intent without modifying the underlying models.

    \item A qualitative case study demonstrating how the same corpus can be reorganized under multiple semantic perspectives.

    \item A simulation-based evaluation of alignment, interaction efficiency, and augmentation strategies.
    
\end{itemize}

\section{Related Work}

\begin{table*}[t]
\centering
\small
\renewcommand{\arraystretch}{0.85}
\caption{Comparison between prior semantic interaction approaches and LLM-augmented semantic steering.}
\label{tab:comp}
\begin{tabular}{p{3.5cm} p{5.6cm} p{5.5cm}}
\toprule
\textbf{Aspect} & \textbf{Prior Approaches} & \textbf{This Work} \\
\midrule
Interpretation of interaction & Encoded implicitly through geometric constraints, feature weighting, or model updates~\cite{endert2012semantic, self2018observation, bian2021deepsi} & Externalized from grouped examples as explicit semantic representations \\
\addlinespace
Representation of semantics & Embedded in model parameters~\cite{bian2021deepsi, self2018observation} or represented as labels or concept descriptors~\cite{el2019semantic, li2022incorporation, oliveira2025creating} & Expressed as natural-language representations at both cluster and document levels \\
\addlinespace
Incorporation mechanism & Requires retraining, metric learning, or model-specific adaptation~\cite{bian2021deepsi, brown2012dis, li2022incorporation} & Uses text augmentation or embedding-level blending without retraining \\
\addlinespace
Model dependence & Often tied to specific models, features, or distance functions~\cite{brown2012dis, li2022incorporation, lin2024imagesi} & Model-agnostic and compatible with black-box embeddings \\
\bottomrule
\end{tabular}
\end{table*}

\hspace*{\parindent}\textbf{Projection-Based Text Visualization.}
Projection-based text visualization is commonly used to explore document collections through low-dimensional views of text embeddings \cite{atzberger2026evaluating}. These projections map text embeddings into 2D spaces using techniques such as UMAP \cite{mcinnes2018umap}, revealing spatial structure that helps analysts examine relationships among documents \cite{atzberger2026evaluating, liu2024visualizing, huang2023va+}. In visual text analysis, projection layouts are often used as interactive workspaces that support exploratory reasoning about semantic structure \cite{bian2021deepsi, sacha2016visual}. However, the organization of a projection is largely determined by the underlying embedding model and dimensionality reduction method, which may not align with semantic relationships that analysts consider meaningful for a given analytic perspective. This limitation has motivated approaches that enable analysts to reshape or steer projection spaces through interaction.

\textbf{Interactive Projection Steering and Semantic Interaction.}
Prior work has explored how user interaction can influence embedding and projection spaces. 
Interactive projection steering interprets user actions---such as grouping, repositioning, or weighting---as signals for updating similarity relationships, feature weights, or model parameters \cite{bradel2014multi, dowling2018sirius, dowling2018bidirectional, sacha2016visual, endert2012semantic, gehrmann2019visual}. Systems such as \textit{Dis-Function} \cite{brown2012dis} and \textit{Andromeda} \cite{self2018observation} learn distance functions or adjust feature weightings based on user feedback, enabling projections that better reflect perceived similarity. 
More broadly, semantic interaction approaches treat user input as implicit constraints that guide metric learning, feature reweighting, or embedding updates \cite{dowling2019interactive, self2018observation, bian2021deepsi, lin2024imagesi, wei2024spaceediting}. While these methods allow analysts to influence projection structure, semantic intent is typically encoded indirectly through geometric or parametric updates. As a result, the relationship between analyst reasoning and representation changes can be difficult to interpret. In addition, many approaches require retraining models or modifying projection algorithms, which limits their applicability to modern black-box embedding models.

\textbf{Explicit Semantic Signals for Projection Steering.}
Several approaches incorporate explicit semantic information into embedding and projection spaces. Li and Zhou \cite{li2022incorporation} let analysts group samples into classes and use these class assignments to train an embedding network with a classification loss, producing more class-consistent projections. \textit{Semantic Concept Spaces} \cite{el2019semantic} enable analysts to define and adjust a hierarchy of concepts and descriptors over a word-embedding space to refine the underlying topic model. More recently, Oliveira et al. \cite{oliveira2025creating} use natural-language prompts that specify a set of candidate categories and zero-shot multimodal LLM classification to derive semantic labels, which are then integrated with data embeddings to influence projection layouts. These methods demonstrate the potential of language- and label-based signals, but they often rely on predefined labels, dataset-wide annotation, or uniform application of generated semantics.

In contrast, our approach derives semantic representations directly from analyst-defined groupings, without predefined labels or global annotation. These representations are expressed in natural language and used as intermediate signals for representation updates. Semantic intent is extended selectively to related documents only when sufficient evidence is present, thereby avoiding uniform application across the dataset. Table~\ref{tab:comp} summarizes the key differences between prior semantic interaction approaches and our LLM-augmented semantic steering framework.

\section{LLM-Augmented Semantic Steering Approach}

We treat low-dimensional projections as semantic workspaces that analysts can reshape to reflect evolving semantic intent. 

\begin{figure*}[t]
    \centering
    \includegraphics[width=\linewidth]{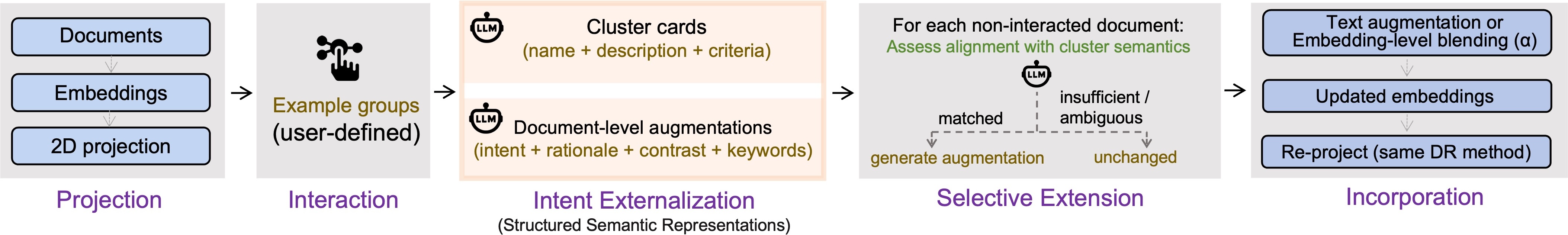}
    \caption{Overview of LLM-augmented semantic steering. Analysts express semantic intent by grouping a small set of documents within the projection. An LLM externalizes this intent as structured cluster- and document-level semantic representations and selectively extends it to related documents. The resulting semantic information is incorporated into document representations via text augmentation or embedding-level blending, and the projection is recomputed from the updated representations.
    }
    \Description{A horizontal pipeline diagram with five stages connected by arrows. From left to right: (1) Projection, showing documents, embeddings, and a 2D layout; (2) Interaction, where a user groups a small set of example documents; (3) Intent Externalization, where an LLM generates cluster-level summaries (name, description, criteria) and document-level semantic augmentations; (4) Selective Extension, where non-interacted documents are compared against the cluster semantics and either receive augmentations if matched or remain unchanged if evidence is insufficient or ambiguous; and (5) Incorporation, where semantic information is integrated through text augmentation or embedding-level blending to produce updated embeddings that are re-projected.
    }
    \label{fig:semantic-pipeline}
\end{figure*}

As illustrated in Figure~\ref{fig:semantic-pipeline}, the approach consists of four stages: \textit{expressing} semantic intent through interaction, \textit{externalizing} the intent as semantic representations, \textit{selectively extending} this intent to related documents, and \textit{incorporating} it into document embeddings to update the projection.

\subsection{Expressing Semantic Intent through Interaction}
Analysts interact with projection spaces to reason about relationships among documents. In our approach, grouping a small number of documents is interpreted as an \textbf{explicit expression of semantic intent}. When analysts group documents, they indicate that these documents are semantically related under the current analytic goal. This relation may reflect shared topics, sentiment, functional roles, or other task-specific criteria that are not fully captured by the original embedding space. Grouping does not assign labels or define fixed categories for the dataset. Instead, grouped documents serve as examples that ground a semantic relation of interest. Interaction is lightweight and selective: analysts typically group only a small subset of documents rather than exhaustively annotating the dataset. By treating interaction as an expression of semantic intent, the approach allows analysts to directly communicate what they consider meaningful within the projection, without specifying similarity metrics or modifying model parameters. 
Analysts can form one or more groups, each containing a few example documents, through direct selection in the projection.

\subsection{Externalizing Semantic Intent as Semantic Representations}
\label{sec:externalizing}

Interactions in the projection space convey semantic intent implicitly. To make this intent explicit and usable in subsequent stages, we externalize it as structured semantic representations using an LLM.

\textbf{Cluster-level semantic descriptions (cluster cards).}  
For each analyst-defined group, the LLM produces a structured cluster card that summarizes the shared meaning of the grouped documents. Each cluster card includes a short cluster name, a concise description of the shared meaning, and brief inclusion and exclusion criteria that characterize what types of documents belong to or differ from the group. The grouping itself is treated as fixed: the LLM does not modify group membership or introduce new groups, but instead articulates the semantic intent implied by the analyst's interaction. These descriptions make the analyst's intent explicit and inspectable, allowing  analysts to verify whether the generated interpretation aligns with their reasoning.

\textbf{Document-level semantic augmentations.}
In addition to cluster-level descriptions, the LLM generates document-level semantic augmentations for each grouped document. These augmentations describe how an individual document exemplifies the shared semantic intent, including a concise statement of the group intent, a document-specific justification, a brief contrast with other analyst-defined groups, and content-specific keywords. While cluster-level descriptions capture the collective meaning of a group, document-level augmentations provide instance-specific connections between individual documents and the articulated intent.

These outputs serve as intermediate language-based representations of semantic intent. They are not used as categorical labels or as final assignments, but as interpretable signals that guide subsequent extension and incorporation into document representations.

\subsection{Selective Extension of Semantic Intent}
In realistic analytic scenarios, analysts interact with only a small subset of documents. To reshape the projection beyond this subset, the expressed semantic intent must be extended to other documents in the collection. Our approach performs this extension in a controlled and selective manner, guided by the semantic representations externalized in the previous stage. 

Cluster cards provide high-level semantic references by describing the intended concept and its boundaries, while document-level augmentations capture how individual documents involved in the interaction instantiate that concept. These representations serve as language-based evidence for assessing semantic relatedness beyond the interacted subset. For each document not involved in the interaction, the LLM compares the document against the existing cluster semantics and determines whether it aligns with one of the analyst-defined groups. 
If so, the document is associated with that cluster and receives a new document-level augmentation of the same form as in Section~\ref{sec:externalizing}, with existing augmentations serving as few-shot examples to guide generation.
If the evidence is weak or the document ambiguously matches multiple clusters, no augmentation is produced and the document is left unchanged.

This design emphasizes selective extension over exhaustive assignment. The goal is not to force every document into one of the expressed semantic groups, but to conservatively extend analyst intent only where sufficient evidence is present, leaving ambiguous or unrelated documents unchanged.

\subsection{Incorporating Semantic Representations into Projection Spaces}

Once semantic intent has been externalized and selectively extended, it must be incorporated into document representations to influence the projection space. Our approach operates at the level of document representations, enabling lightweight and model-agnostic semantic steering. Document-level semantic augmentations are expressed as concise natural-language text that complements the original document. These augmentations encode the analyst's semantic intent and can be incorporated in several ways.

\textbf{Text-based augmentation.} 
Semantic augmentations can be appended or prepended to the original document text before embedding. Optional tags can be used to mark the original content and the augmentation separately, enabling additional variants of this strategy. This approach allows the embedding model to integrate both the original content and the expressed semantic intent when computing document representations.

\textbf{Embedding-level blending.} Alternatively, semantic information can be incorporated at the embedding level by combining the original document embedding with the embedding of the semantic augmentation.
Let $E_{\text{base}}$ denote the original document embedding and $E_{\text{aug}}$ the embedding of the semantic augmentation. The updated embedding is computed as $E' = (1-\alpha)E_{\text{base}} + \alpha E_{\text{aug}}$, where $\alpha \in [0,1]$ controls the strength of semantic steering. Smaller values of $\alpha$ preserve more of the original structure, while larger values emphasize intent-aligned organization.

After updating document representations, the projection is recomputed using the same dimensionality reduction method as in the baseline. As a result, changes in the projection reflect shifts in semantic emphasis induced by analyst intent, rather than changes to the projection algorithm itself. Because semantic representations remain explicit and inspectable, analysts can reason about how and why the projection changes as intent is expressed and refined.
Prompt templates, output schemas, and representative examples are provided in the supplementary material.

\section{Case Study: Semantic Steering under Multiple Analytic Perspectives}

We present a qualitative case study illustrating how LLM-augmented semantic steering reorganizes projection spaces under different semantic perspectives through lightweight interaction.

\subsection{Dataset and Analytic Context}
We use a subset of Amazon product reviews comprising 120 documents across four product categories---PC, Drugstore, Sports, and Apparel---with 30 reviews per category \cite{keung2020multilingual}. Each category contains an equal number of positive and negative reviews (15 each), yielding a balanced distribution across both product type and sentiment. 
This dataset supports multiple analytic perspectives, including product-oriented views based on topical distinctions and sentiment-oriented views based on evaluative language.
The baseline embedding and projection follow the setup described in Section~\ref{sec:evaluation_setup}, and semantic augmentations are incorporated via embedding-level blending with $\alpha = 0.75$.

\subsection{Baseline Projection: An Underspecified Semantic Workspace}
We begin with an unsteered projection based solely on the original review text. In this baseline layout (Figure~\ref{fig:teaser}a), reviews from different product categories and sentiment polarities are intermingled, and local neighborhoods primarily reflect general linguistic similarity, emphasizing neither product category nor sentiment structure. This baseline projection is not incorrect, but underspecified with respect to any particular analytic perspective, motivating the need for mechanisms that allow analysts to reshape projection spaces according to semantic intent.

\subsection{Steering toward a Category Perspective}
We first consider a category-oriented perspective, in which the analyst aims to understand how reviews relate to different types of products.

\textbf{Expressing the Category Perspective.} The analyst selects a small number of reviews that exemplify several product-related concepts, such as clothing, electronics, and health-related products. These reviews are grouped through interaction in the projection space (Figure~\ref{fig:teaser}b), forming example-based groups that express the intended semantic organization.

\textbf{Externalized Category Semantics.}
From these grouped examples, the framework externalizes structured semantic representations that articulate shared product-related characteristics (Figure~\ref{fig:teaser}d). At the cluster level, these representations describe recurring aspects of products, such as fit and comfort for clothing, effectiveness and side effects for health-related items, or functionality and reliability for electronic devices.

For example, a clothing-related cluster may be summarized as \textit{``reviews of garments emphasizing fit, sizing, body shape, and comfort, including references to fabric, coverage, and wearability.''} At the document level, semantic augmentations further connect these cluster-level descriptions to individual reviews. A review in this cluster may be characterized as \textit{``describing tight fit and uncomfortable material, highlighting how fabric and sizing affect wearability for different body types.''} These representations make the analyst's semantic intent explicit and inspectable.

\textbf{Category-Oriented Projection and Interpretation.}
After incorporating these semantic augmentations into document embeddings, the projection is updated. In the resulting layout (Figure~\ref{fig:teaser}c), reviews associated with similar product types form more coherent regions than in the baseline projection, and distinctions between product-related regions become easier to perceive. For example, clothing-related reviews cluster around discussions of fit and comfort, while electronics-related reviews emphasize functionality and performance.
This reflects a shift in semantic emphasis driven by the grouped examples.
From an analytic perspective, the updated projection enables the analyst to more easily identify regions associated with particular product types and to compare similarities and differences within and across these regions.

\subsection{Steering toward a Sentiment Perspective}
We next consider a sentiment-oriented perspective, in which the analyst focuses on evaluative language.

\textbf{Expressing the Sentiment Perspective.}
Using the same baseline projection setup, the analyst selects a small number of reviews that exemplify positive and negative sentiment (Figure~\ref{fig:teaser}b). These reviews are grouped through interaction to express an evaluative semantic concept, in which documents are considered similar based on their expressed opinions.

\textbf{Externalized Sentiment Semantics.}
From these grouped examples, the framework generates semantic representations that capture patterns of evaluative language. At the cluster level, these representations describe recurring forms of positive and negative sentiment, such as strong dissatisfaction due to product failure or overall satisfaction with minor caveats (Figure~\ref{fig:teaser}d).

For example, a negative-sentiment cluster may be summarized as \textit{``reviews expressing dissatisfaction due to product failure, poor experience, or unmet expectations, often using strongly evaluative language.''} At the document level, augmentations link these representations to individual reviews by highlighting specific expressions of frustration or disappointment. A review may be characterized as \textit{``emphasizing discomfort and dissatisfaction with product quality, framing the experience as a clear failure rather than a minor inconvenience.''} Compared to the category-oriented representations, these representations focus on affective and evaluative aspects of the text instead of product attributes or functionality.

\textbf{Sentiment-Oriented Projection and Interpretation.}
After incorporating these semantic augmentations, the projection is updated. In the resulting layout (Figure~\ref{fig:teaser}c), reviews with similar sentiment become more spatially coherent, even when they refer to different product categories. The projection shifts from topical similarity toward evaluative similarity, making regions of positive and negative sentiment more clearly distinguishable. From an analytic perspective, this view enables the analyst to inspect how sentiment is distributed across product types, identify regions dominated by strong approval or dissatisfaction, and examine borderline or mixed cases.

\subsection{Reorganization across Semantic Perspectives}
Comparing the category-oriented and sentiment-oriented analyses highlights a key property of semantic steering: the same dataset can be reorganized under different semantic perspectives without modifying the underlying embedding model or projection method. Each perspective emphasizes different semantic dimensions (e.g., product attributes vs. evaluative language), leading to distinct projection structures.

Notably, the same document may be interpreted differently across semantic perspectives. For example, a review describing tight fit and uncomfortable material may be associated with negative sentiment in one view, while in a category-oriented view, it is grouped with clothing-related discussions of fit and wearability. This illustrates how semantic steering reveals different aspects of the same content depending on analytic goals.

\section{Evaluation}
We evaluate semantic steering as a computational mechanism for incorporating analyst-expressed semantic intent into projection spaces. Our goal is to assess whether lightweight, example-based interaction can reliably reshape projection structures to align with a target semantic organization, while remaining efficient and controllable.

To enable systematic and reproducible evaluation, we design simulation-based experiments in which a small set of documents serves as input examples for interaction. Predefined semantic groupings are used only as reference structures for evaluation. This setup allows us to isolate the effects of semantic steering under controlled conditions without requiring exhaustive manual interaction.

\subsection{Evaluation Setup}
\label{sec:evaluation_setup}
\hspace*{\parindent}\textbf{Dataset and reference grouping.} We conduct experiments on a corpus of 112 papers sampled from the IEEE VIS 2022 and 2023 proceedings, represented by titles and abstracts \cite{ieeevis}. For evaluation, we derive reference semantic groups based on conference session categories, further merged into four high-level groups: \textit{Immersive and Interactive Visualization} (17 papers), \textit{Domain-Specific Visualization Applications} (34 papers), \textit{Educational, Public, and Storytelling Visualization} (30 papers), and \textit{Machine Learning and AI in Visualization} (31 papers). These reference groups are used solely for evaluation and are not provided to the LLM or used in steering.

\textbf{Baseline embeddings and projection.} Documents are embedded using the \texttt{text-embedding-3-small} model \cite{openai_api} and projected into 2D using UMAP with cosine distance (\texttt{n\_neighbors = 15}, \texttt{min\_dist = 0.1}). This unsteered projection serves as the baseline.

\textbf{Simulated interaction.} To study steering behavior under controlled and reproducible conditions, we simulate analyst interaction by sampling a small number of documents from each reference group and treating them as example-based groups. Unless otherwise specified, five documents per group are randomly selected. This setup models a scenario in which an analyst provides a small set of examples to express semantic intent.
All LLM-based components use GPT-5.1 via the OpenAI API \cite{openai_api} at temperature 0.

\textbf{Evaluation metrics.} Projection quality is assessed using two complementary measures. \textit{Global alignment} is measured by a scaled silhouette score \cite{lin2024imagesi} ($\mathrm{Sil} = 2s$, where $s$ is the standard silhouette score \cite{rousseeuw1987silhouettes}) computed with respect to the reference semantic grouping. Following prior sensemaking-oriented evaluation, $\mathrm{Sil} \approx 1$ (i.e., $s \approx 0.5$) is treated as the ideal value, reflecting well-separated but not overly compact semantic groups \cite{lin2024imagesi, han2023explainable}. \textit{Local alignment} is measured by neighborhood consistency (NC), defined as the average fraction of same-group documents among each point's $k$ nearest neighbors in the 2D projection ($k = 10$). Both measures are reported as changes relative to the unsteered baseline ($\Delta \mathrm{Sil}$, $\Delta \mathrm{NC}$), with positive values indicating improved alignment.

\textbf{Random augmentation control.} As a non-semantic control, we replace semantic augmentations with length-matched random text sampled from the same document. This allows us to distinguish the effect of semantic content from generic textual perturbation.

\subsection{RQ1: Does semantic steering increase
alignment with a target semantic structure?}

We first examine whether semantic steering improves alignment between projection layouts and a target semantic structure, using the reference grouping as the evaluation target and the unsteered projection as the baseline.

Table~\ref{tab:text_aug} summarizes the effects of text-based augmentation strategies. Across all semantic variants, steering consistently improves both $\Delta \mathrm{Sil}$ and $\Delta \mathrm{NC}$, indicating stronger global separation and more coherent local neighborhoods with respect to the reference semantic grouping. These improvements are consistent across multiple projection initializations. In contrast, random augmentation controls do not produce comparable gains and often degrade projection quality. This indicates that the observed improvements arise from semantically meaningful augmentation, not from generic textual perturbation. Among the semantic variants, augmentation-only yields the strongest improvements in global alignment, suggesting that semantic signals alone can strongly influence projection structure. However, strategies that retain the original document content (e.g., prepend or append) provide more balanced behavior by improving alignment while preserving aspects of the baseline structure. Across these strategies, we observe comparable improvements regardless of whether augmentations are appended, prepended, or tagged, indicating robustness to surface-level design choices.

\begin{table}[t]
\centering

\caption{Effects of text-based augmentation strategies on projection alignment. Values are mean $\pm$ std over five UMAP initializations relative to the baseline (Sil $= 0.21 \pm 0.04$, 
NC $= 0.55 \pm 0.02$). Higher $\Delta \mathrm{Sil}$ and $\Delta \mathrm{NC}$ indicate stronger alignment. Semantic augmentation improves alignment, while random augmentation degrades performance.}
\label{tab:text_aug}
\small
\begin{tabular}{llcc}
\toprule
\textbf{Strategy} & \textbf{Formulation} & \textbf{$\Delta$Sil $\uparrow$} & \textbf{$\Delta$NC $\uparrow$} \\
\midrule
\multicolumn{4}{l}{\textbf{Semantic augmentation}} \\
Append            & DOC + AUG              & 0.27$\pm$0.04 & 0.07$\pm$0.03 \\
Prepend           & AUG + DOC              & 0.29$\pm$0.02 & 0.09$\pm$0.02 \\
Tagged append    & \texttt{<ORG>} DOC + AUG    & 0.27$\pm$0.01 & 0.08$\pm$0.02 \\
Tagged prepend   & AUG + \texttt{<ORG>} DOC    & 0.26$\pm$0.06 & 0.08$\pm$0.03 \\
Augmentation only & AUG                    & \textbf{0.47$\pm$0.05} & \textbf{0.10$\pm$0.02} \\
\midrule
\multicolumn{4}{l}{\textbf{Random augmentation control}} \\
Append            & DOC + RAND             & -0.09$\pm$0.01 & -0.05$\pm$0.02 \\
Prepend           & RAND + DOC             & -0.12$\pm$0.03 & -0.07$\pm$0.02 \\
Tagged append    & \texttt{<ORG>} DOC + RAND   & -0.10$\pm$0.03 & -0.05$\pm$0.02 \\
Tagged prepend   & RAND + \texttt{<ORG>} DOC   & -0.15$\pm$0.03 & -0.10$\pm$0.03 \\
Augmentation only & RAND                   & \textbf{-0.23$\pm$0.04} & \textbf{-0.15$\pm$0.02} \\
\bottomrule
\end{tabular}

\vspace{2pt}
\footnotesize\textit{Note:} DOC = original document; AUG = semantic augmentation; RAND = random text augmentation; \texttt{<ORG>} marks original content in tagged variants.

\end{table}

\begin{figure*}[t]
    \centering
    \includegraphics[width=\linewidth]{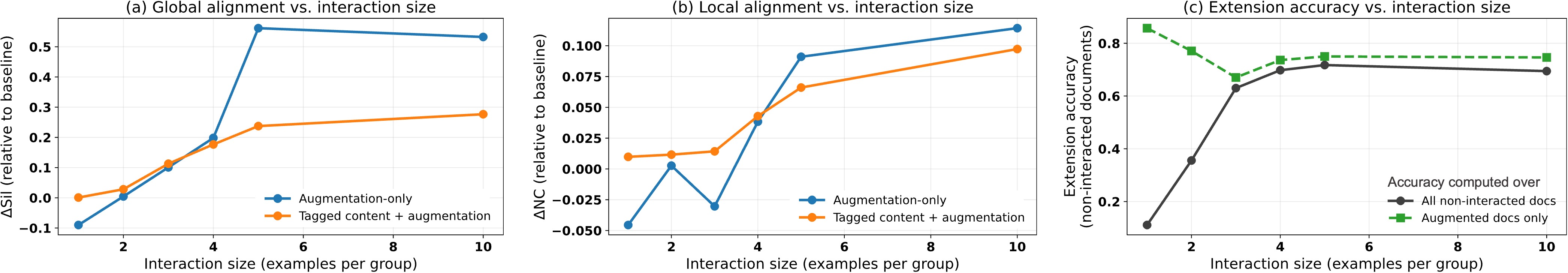}
    \caption{
    Interaction efficiency and selective extension in semantic steering.
    (a--b) Global ($\Delta \mathrm{Sil}$) and local ($\Delta \mathrm{NC}$) alignment improve with small interaction sizes and plateau thereafter, indicating that a few examples are sufficient to capture semantic intent.
    (c) Extension accuracy on documents not involved in the interaction increases at small interaction sizes and then stabilizes, with higher accuracy for documents that receive semantic augmentations.
    }
    \Description{Three line charts showing the effect of interaction size on semantic steering. The x-axis represents the number of example documents per group. Chart (a) shows changes in global alignment for two strategies, both increasing with interaction size and then leveling off. Chart (b) shows changes in local alignment, with both strategies increasing and converging at higher interaction sizes. Chart (c) shows extension accuracy for non-interacted documents, with one curve for all documents and another for those receiving semantic augmentations; accuracy increases initially and then stabilizes, with consistently higher values for augmented documents.
    }
    \label{fig:interaction-propagation}
\end{figure*}

\subsection{RQ2: How much interaction is needed for effective steering?}
We next examine how the amount of interaction influences the effectiveness of semantic steering. 
To study this, we vary the number of example documents sampled per reference group as input to semantic steering.

Figures~\ref{fig:interaction-propagation}a and~\ref{fig:interaction-propagation}b show changes in $\Delta \mathrm{Sil}$ and $\Delta \mathrm{NC}$ as interaction effort increases,
reporting two representative strategies from Table~\ref{tab:text_aug} (augmentation-only and tagged content + augmentation); other strategies exhibit similar patterns. Gains in both global and local alignment appear with a small number of examples per group and begin to plateau around five examples, beyond which additional interaction yields only marginal returns.
This suggests that once semantic intent has been sufficiently captured, further interaction provides limited new information. Overall, these results indicate that effective semantic steering requires only minimal interaction, supporting it as a practical, lightweight mechanism for intent-driven analysis.

\subsection{RQ3: Can semantic information be extended beyond interacted documents?}

Using the same interaction-size setup, we now turn to how far the expressed intent propagates beyond the interacted documents, and whether it can be reliably extended to the rest of the collection.

Figure~\ref{fig:interaction-propagation}c reports extension accuracy on documents not involved in the interaction as interaction size increases. Accuracy is computed with respect to the reference grouping, both over all non-interacted documents and over those that received semantic augmentations. Extension accuracy improves substantially as interaction increases from one to five examples per group, and stabilizes thereafter. Accuracy is consistently higher when evaluated on augmented documents, indicating that the extension mechanism selectively generates augmentations for documents with sufficient semantic evidence; weakly related or ambiguous documents are left unchanged. This selectivity is also reflected in coverage: the proportion of non-interacted documents that receive augmentations grows with interaction size, from roughly $13\%$ at one example per group to about $96\%$ at five.
These findings show that semantic steering selectively extends intent beyond the interacted subset, rather than exhaustively as in label propagation or classification.

\begin{table}[t]

\centering
\caption{Effects of embedding-level blending on projection quality. Values are mean $\pm$ std over five UMAP initializations relative to the unsteered baseline. As $\alpha$ increases, global alignment ($\Delta \mathrm{Sil}$) improves while local alignment ($\Delta \mathrm{NC}$) increases initially and then stabilizes.}
\label{tab:blending}
\small
\begin{tabular}{clcc}
\toprule
\textbf{$\alpha$} & \textbf{Formulation} & \textbf{$\Delta$Sil $\uparrow$} & \textbf{$\Delta$NC $\uparrow$} \\
\midrule
0.00 & $E = E_{\text{base}}$                              & 0.00$\pm$0.00 & 0.00$\pm$0.00 \\
0.25 & $E = 0.75E_{\text{base}} + 0.25E_{\text{aug}}$    & 0.13$\pm$0.03 & 0.07$\pm$0.02 \\
0.50 & $E = 0.50E_{\text{base}} + 0.50E_{\text{aug}}$    & 0.32$\pm$0.08 & 0.10$\pm$0.03 \\
0.75 & $E = 0.25E_{\text{base}} + 0.75E_{\text{aug}}$    & 0.40$\pm$0.08 & 0.10$\pm$0.03 \\
1.00 & $E = E_{\text{aug}}$                               & \textbf{0.47$\pm$0.04} & 0.09$\pm$0.02 \\
\bottomrule
\end{tabular}

\end{table}

\begin{figure*}[t]
    \centering
    \includegraphics[width=\linewidth]{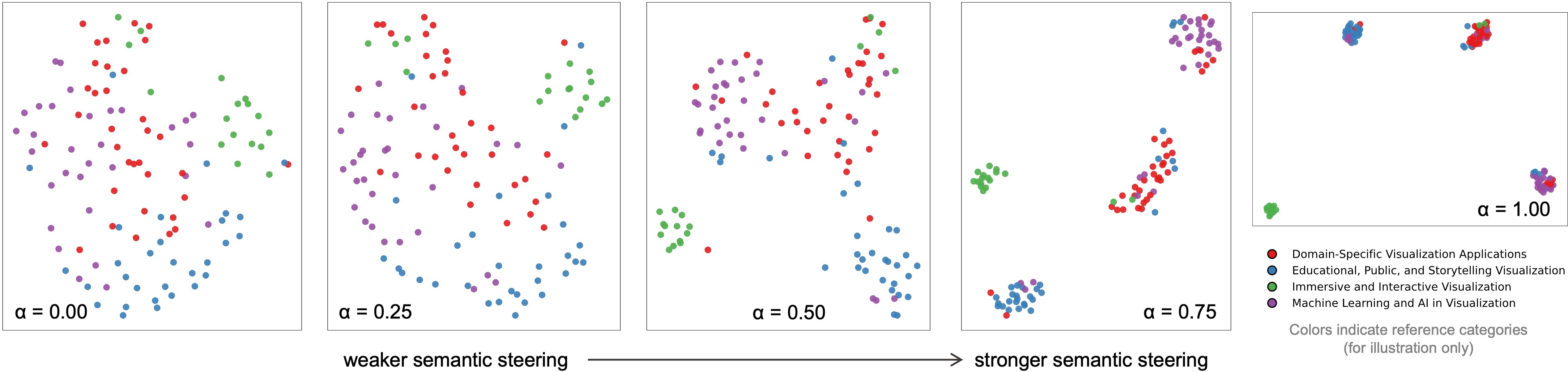}
    \caption{
    Progressive semantic steering via embedding-level blending on the IEEE VIS corpus. As the blending weight $\alpha$ increases, projections transition smoothly from the unsteered baseline ($\alpha$ = 0) toward more semantically structured layouts. Global separation increases with respect to the reference semantic grouping, while local neighborhood structure remains relatively stable, demonstrating continuous and controllable steering.
    }
    \Description{A sequence of five 2D scatter plots arranged from left to right, corresponding to increasing blending weights alpha from 0.00 to 1.00. Points are colored by four reference categories. At low alpha, points of different colors are intermixed with no clear group structure. As alpha increases, points of the same color progressively form more compact and separated clusters. An arrow below the plots indicates the progression from weaker to stronger semantic steering.
    }
    \label{fig:progressive-steering}
\end{figure*}

\subsection{RQ4: How do augmentation strategies affect steering behavior?}
Having established that semantic steering improves alignment and operates effectively with limited interaction, we examine how different augmentation strategies influence steering behavior.

Table~\ref{tab:blending} summarizes the effects of embedding-level blending, where the blending weight $\alpha$ controls interpolation between the base and augmentation embeddings. As $\alpha$ increases, $\Delta \mathrm{Sil}$ consistently improves, indicating progressively stronger global separation with respect to the reference semantic grouping. $\Delta \mathrm{NC}$ increases initially and then stabilizes, suggesting that local neighborhood structure is preserved even as global organization becomes better aligned. Unlike text-based augmentation strategies in Table~\ref{tab:text_aug}, which apply semantic information at a fixed strength, blending provides a continuous mechanism for controlling steering strength. Even moderate values of $\alpha$ yield substantial improvements, and higher values further emphasize the target semantic structure. Figure~\ref{fig:progressive-steering} illustrates this behavior qualitatively: as $\alpha$ increases, the projection transitions smoothly from the unsteered baseline toward a more structured layout, demonstrating continuous control over steering strength. In our experiments, intermediate values of $\alpha$ (e.g., 0.5--0.75) provide a balance between improving global alignment and preserving baseline structure. Overall, embedding-level blending offers a simple and effective mechanism for controlling semantic steering.

\section{Discussion}
\hspace*{\parindent}\textbf{Semantic Interaction as Intent Externalization.} This work reframes semantic interaction in projection-based text analysis as a process of intent externalization rather than parameter manipulation. Prior approaches typically interpret analyst interaction implicitly, translating spatial actions into geometric constraints or model updates \cite{dowling2019interactive, self2018observation, endert2012semantic, bian2021deepsi}. In contrast, our approach treats interaction as a semantic statement: analysts express what is meaningful to them through grouping, and this intent is externalized explicitly in natural language. By using language as an intermediate representation, our approach makes analyst-expressed semantic intent inspectable and revisable while remaining model-agnostic. This shifts interaction from indirect manipulation of model behavior toward \textit{more direct expression of analytic reasoning}.

\textbf{Projections as Intent-Dependent Semantic Workspaces.}
Our findings support viewing projections not as fixed approximations of an underlying ``true'' semantic structure, but as \emph{intent-dependent semantic workspaces}. Different analytic goals may emphasize different semantic dimensions, and corresponding changes in projection structure reflect shifts in semantic emphasis. 
Evaluating projections solely in terms of fidelity to the original embedding space therefore captures only part of their analytic role \cite{jeon2025stop, atzberger2026evaluating}. Semantic steering complements traditional dimensionality reduction by prioritizing alignment with analytic intent while preserving structural properties of the embedding space.

\textbf{Transparency, Trust, and Analyst Control.} 
Extending semantic intent beyond documents directly involved in the interaction introduces uncertainty regarding how far and how accurately such intent propagates. To address this, our approach emphasizes transparency and supports analyst control by externalizing semantic intent as explicit natural-language representations that can be inspected and revised by the analyst. This explicitness can foster analyst trust in two ways. First, it helps analysts understand why a projection changes as intent is expressed or revised, linking spatial reorganization to articulated semantic meaning. Second, the selective and conservative design of intent extension is intended to avoid silent over-extension, so that cases where extended semantics diverge from analyst expectations remain visible and inspectable. In this setting, trust arises not from guarantees of correctness, but from the ability to \textit{inspect}, \textit{monitor}, and iteratively refine how semantic intent is incorporated \cite{vaithilingam2025semantic}.

\textbf{Limitations and Future Directions.} 
While the results demonstrate the potential of LLM-augmented semantic steering, several limitations suggest directions for future work. First, the approach relies on the interpretive capability of LLMs, which may introduce variability or occasional misalignment with analyst intent \cite{wang2024survey}. Although externalizing semantic interpretations makes such behavior visible, improving robustness and consistency remains an important challenge. Second, scalability remains an open question. Generating and extending semantic representations incurs computational cost, particularly for large document collections. Future work could explore incremental updates, caching strategies, or hybrid approaches that reduce reliance on repeated LLM calls while preserving interpretability. Finally, our current interaction design focuses on grouping as the primary means of intent expression. Richer forms of semantic interaction, such as negative intent, hierarchical relationships, or evolving analytic goals, represent a promising extension of this design space \cite{endert2012semantic}. User studies examining how semantic steering integrates with diverse analytic workflows and downstream tasks would further clarify its practical impact \cite{endert2012semantics}.

\section{Conclusion}
We introduce LLM-augmented semantic steering for reshaping projection spaces of text embeddings. Analysts express semantic intent by grouping a small set of example documents. This intent is externalized as structured semantic representations, selectively extended to related documents, and incorporated into document representations without retraining the underlying models.
Through a case study and simulation-based evaluation, we show that semantic steering improves alignment with target semantic structures, requires only minimal interaction, and supports selective extension beyond documents directly involved in the interaction. Embedding-level blending enables continuous and controllable steering.
More broadly, this work highlights the potential of language-mediated interaction for integrating human semantic reasoning with representation learning in interactive visual analysis systems.

\begin{acks}
This research was supported by industry, government, and institute members of the NSF SHREC Center, which was founded in the IUCRC program of the National Science Foundation.
\end{acks}

\bibliographystyle{ACM-Reference-Format}
\bibliography{references}


\end{document}